\documentclass[12pt]{article}
\usepackage{amscd,amsmath,amssymb,amstext,amsthm,exscale,latexsym}
\usepackage{graphicx,floatflt}
\newcommand {\g }   {\gamma}       
\newcommand {\dl}   {\delta}       
        
\newcommand {\ve}   {\varepsilon}

\newcommand {\vf }  {\varphi}      
         \newcommand {\om}  {\omega}
      \newcommand {\Om}  {\Omega}
\newcommand {\Th}   {\Theta}       
\newcommand {\pl}   {\partial}

\newcommand   {\const}{{\sf\,const}}     
         
\renewcommand {\sin}{{\sf\,sin\,}}       \renewcommand {\cos}{{\sf\,cos\,}}

\newcommand {\MC}  {{\mathbb C}}

\newcommand {\MM}  {{\mathbb M}}   
\newcommand {\MO}  {{\mathbb O}}   
   \newcommand {\MR}  {{\mathbb R}}
\newcommand {\MS}  {{\mathbb S}}

   \newcommand {\MZ}  {{\mathbb Z}}

\begin{document}
\title     {Chern--Simons term in the geometric theory of defects}
\author    {M. O. Katanaev
            \thanks{E-mail: katanaev@mi.ras.ru}\\
            \sl Steklov mathematical institute,\\
            \sl ul.~Gubkina, 8, Moscow, 119991, Russia}
\date      {19 May 2017}
\maketitle
\begin{abstract}
The Chern--Simons term is used in the geometric theory of defects. The
equilibrium equations with $\dl$-function source are explicitly solved with
respect to the $\MS\MO(3)$ connection. This solution describes one straight
linear disclination and corresponds to the new kind of geometrical defect: it is
the defect in the connection but not the metric which is the flat Euclidean
metric. This is the first example of a disclination described within the
geometric theory of defects. The corresponding angular rotation field is
computed.
\end{abstract}
\section{Introduction}
Ideal crystals are absent in nature, and most of their physical properties, such
as plasticity, melting, growth, etc., are defined by defects of the crystalline
structure. Therefore, a study of defects is a topical scientific question of
importance for applications in the first place. At present, a fundamental
continuous theory of defects is absent in spite of existence of dozens of
monographs and thousands of articles.

One of the most promising approaches to the theory of defects is based on
Riemann--Cartan geometry, which involves nontrivial metric and torsion.
In this approach, a crystal is considered as a continuous elastic medium with
a spin structure. If the displacement vector field is a smooth function, then
there are only elastic stresses corresponding to diffeomorphisms of the
Euclidean space. If the displacement vector field has discontinuities, then
we are saying that there are defects in the elastic structure. Defects in the
elastic structure are called dislocations and lead to the appearance
of nontrivial geometry. Precisely, they correspond to a nonzero torsion tensor,
equal to the surface density of the Burgers vector. Defects in the spin
structure are called disclinations. They correspond to nonzero curvature tensor,
curvature tensor being the surface density of the Frank vector.

The idea to relate torsion to dislocations appeared in the 1950s [1--4].
\nocite{Kondo52,Nye53,BiBuSm55,Kroner58}
The review and earlier references can be found in the book \cite{Kleine08}.

In the geometric approach to the theory of defects [6--8], we discuss
\nocite{KatVol92,KatVol99,Katana05}
a model which is different from others in two respects. Firstly, we do not have
the displacement vector field and rotational vector field as independent
variables because, in general, they are not continuous. Instead, the triad field
and $\MS\MO(3)$ connection are considered as the only independent variables. If
defects are absent, then the triad and $\MS\MO(3)$ connection reduce to partial
derivatives of the displacement and rotational angle vector fields. In this
case, the latter can be reconstructed. Secondly, the set of equilibrium
equations is different. We proposed purely geometric set which coincides with
that of Euclidean three dimensional gravity with torsion. The nonlinear
elasticity equations and principal chiral $\MS\MO(3)$ model for the spin
structure enter the model through the elastic and Lorentz gauge conditions
[8--10]
\nocite{Katana03,Katana04,Katana05} which allow us to reconstruct the
displacement and rotational angle vector fields in the absence of dislocations
in full agreement with classical models.

The advantage of the geometric theory of defects is that it allows one to
describe single defects as well as their continuous distributions.

In the present paper, we describe static distribution of linear disclinations
within the geometric theory of defects using the Chern--Simons term. It seems to
be well suited for this problem. The case of one linear straight disclination is
considered in detail.
\section{Disclinations in the geometric theory of defects}
Let us consider a three dimensional continuous medium described by a
Riemann--Cartan manifold equipped with a spin structure. We use triad field
$e_\mu{}^i$ and $\MS\MO(3)$ connection $\om_\mu{}^{ij}=-\om_\mu{}^{ji}$, where
Greek letters $\mu=1,2,3$ and Latin ones $i,j=1,2,3$ denote world and tangent
indices, respectively, as basic independent  variables. To describe
disclinations, we assume that topology of the manifold is a trivial $\MR^3$
topology, metric $g_{\mu\nu}:=e_\mu{}^i e_\nu{}^j\dl_{ij}=\dl_{\mu\nu}$ is an
ordinary flat Euclidean metric, but connection is nontrivial and may have
singularities on isolated subsets of measure zero corresponding to single
disclinations.

\begin{figure}[hbt]
\hfill\includegraphics[width=.35\textwidth]{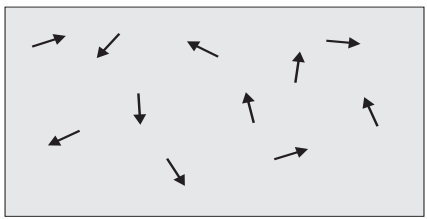}
\hfill {}
\centering\caption{In the continuous approximation, the spin structure of a ferromagnet is described by unit vector field $n(x)$.}
\label{ferromagnet}
\end{figure}

\begin{figure}
\hfill\includegraphics[width=.75\textwidth]{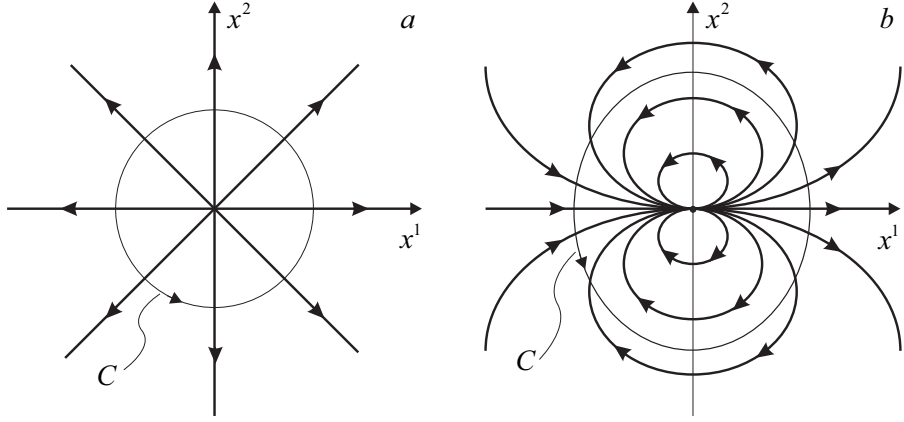}
\hfill {}
\centering\caption{Distribution of unit vector field in the $x,y$ plane for
straight linear disclinations parallel to the $x^3$ axis, for $|\Theta|=2\pi$
(\textit{a}) and $|\Theta|=4\pi$ (\textit{b}).}
\label{fdiscl}
\end{figure}

In the continuous approximation, the spin structure is described by unit vector
field $n^i(x)$ $(n^in_i=1)$ shown in Fig.\ref{ferromagnet}. For example, it
may be a ferromagnet, where $n^i$ describes distribution of magnetic moments.
The unit vector field can be described as follows. We fix some direction in the
medium $n_0^i$. Then the field $n^i(x)$ at a point $x$ can be uniquely defined
by the field $\theta^{ij}(x)=-\theta^{ji}(x) =\frac12\ve^{ijk} \theta_k$,  where
$\ve^{ijk}$ is the totally antisymmetric tensor  and $\theta_k$ is a covector
directed along the rotation axis, its length being the rotation angle. Here and
in what follows, Latin tangent indices are raised and lowered with the help of
the flat Euclidean metric $\dl_{ij}$. We call this field the spin structure of
the medium
$$
  n^i=n_0^j S_j{}^i(\theta),
$$
where $S_j{}^i\in\MS\MO(3)$ is the rotation matrix corresponding to
$\theta^{ij}$ parameterized as (see, e.g., \cite{Katana04})
\begin{equation}                                                  \label{elsogr}
  S_i{}^j=(e^{(\theta\ve)})_i{}^j=\cos\theta\,\dl_i^j
  +\frac{(\theta\ve)_i{}^j}\theta\sin\theta
  +\frac{\theta_i\theta^j}{\theta^2}(1-\cos\theta)\qquad \in\MS\MO(3)\, ,
\end{equation}
where $(\theta\ve)_i{}^j:=\theta_k\ve^k{}_i{}^j$ and
$\theta:=\sqrt{\theta^i\theta_i}$.
If the unit vector field is continuous then there are no defects in the spin
structure which are called disclinations. The simplest examples of linear
disclinations are shown in Fig.~\ref{fdiscl}, where the discontinuity occurs on
a line parallel to the $x^3$ axis and the vector field $n$ lies in the
perpendicular plane $(x^1,x^2)$.

A linear disclination is characterized by the Frank vector
\begin{equation}                                                  \label{etheta}
  \Th_i:=\frac12\ve_{ijk}\Th^{jk},
\end{equation}
where
\begin{equation}                                                  \label{eomega}
  \Th^{ij}:=\oint_Cdx^\mu\pl_\mu\theta^{ij},
\end{equation}
and the integral is taken along closed contour $C$ surrounding the disclination
axis. The length of the Frank vector is equal to the total angle of rotation of
the field $n^i$ as it goes around the disclination. In general, the vector field
$n^i$ defines a map of the Euclidean space to a sphere
$n:~\MR^3\rightarrow\MS^2$. For linear disclinations parallel to the $x^3$ axis,
this map is restricted to a map of the plane $\MR^2$ to a circle $\MS^1$. In
this case, the total rotation angle must obviously be a multiple of $2\pi$. In
the presence of disclinations, the rotational angle field $\theta^{ij}(x)$ is no
longer smooth, and we must make some cuts for a given distribution of
disclinations and impose appropriate boundary conditions in order to define
$\theta^{ij}(x)$.

The previous discussion refers to an isolated disclination and can be naturally
extended to a situation in which there is a countable number of isolated
disclinations, each of them contained in an open neighborhood containing no
other disclinations. If there is a continuous distribution of disclinations the
rotational vector field $\theta^{ij}$ does not exist. In order to describe this
situation it is convenient to introduce a new structure: the $\MS\MO(3)$
connection $\om_\mu{}^{ij}(x)$ which is continuous on the cuts. In this manner,
the geometric theory of defects describes both single defects as well as their
continuous distribution, in which the idea of isolated dislocations is replaced
by the notion of \textit{curvature}. Disclinations are said to be absent if and
only if the curvature of $\MS\MO(3)$ connection vanishes,
\begin{equation}                                                  \label{ubvgfr}
  R_{\mu\nu j}{}^i=\pl_\mu \om_{\nu j}{}^i-\om_{\mu j}{}^k
  \om_{\nu k}{}^i-(\mu\leftrightarrow\nu)=0,
\end{equation}
or in the language of forms,
\begin{equation}                                                  \label{ucvdfr}
  R_i{}^j=\frac12dx^\mu\wedge dx^\nu R_{\mu\nu i}{}^j=d\om_i{}^j-\om_i{}^k\wedge\om_k{}^j=0.
\end{equation}
It will be shown that for an isolated disclination,
$R^{ij}\sim\delta(\g)\ve^{ijk}m_k$, where $m_k$ is the tangent vector to the
disclination line $\g$.

In the absence of disclinations, the $\MS\MO(3)$ connection is a pure gauge, and
there is a rotational angle field $\theta^{ij}$ such that
$\om_\mu{}^{ij}=\pl_\mu\theta^{ij}$. On the contrary, in the presence of
disclinations curvature does not vanish, the connection $\om_\mu{}^{ij}$ is not
a pure gauge and the rotational angle field is not globally defined throughout
$\MM$.

In general, the geometric theory of defects describes two types of defects:
dislocations and disclinations which are defects of the elastic medium itself
and the spin structure, respectively. This general situation corresponds to
nontrivial curvature and torsion. These notions have physical interpretation as
surface densities of the Frank vector (curvature) and surface density of Burgers
vector (torsion) \cite{KatVol92,Katana05}.

We consider the case when dislocations and elastic deformations are absent. Then
the elastic media is the Euclidean space $\MR^3$ with Euclidean metric
$\dl_{ij}$ and zero torsion. Disclinations in the spin structure are described
by $\MS\MO(3)$ connection with nontrivial curvature.
\section{Action for disclinations}
In three dimensions, the $\MS\MO(3)$ connection can be parameterized in the
following way
\begin{equation}                                                  \label{uxnbcg}
  \om_\mu{}^{ij}=\om_{\mu k}\ve^{kij},\qquad\om_{\mu k}:=\frac12\om_\mu{}^{ij}
  \ve_{ijk}.
\end{equation}
Then, curvature tensor takes the form
\begin{equation}                                                  \label{ushtde}
  R_{\mu\nu k}:=R_{\mu\nu}{}^{ij}\ve_{ijk} =2(\pl_\mu\om_{\nu k}
  -\pl_\nu\om_{\mu k}+\om_\mu{}^i\om_\nu{}^j\ve_{ijk}),
\end{equation}
The action is defined by the Chern-Simons form for the connection field of
the continuous medium given by
\begin{equation}                                                      \label{CS}
  \mathcal{E}[\omega]= \int_{\MR^3}\left(\frac12\omega^i\wedge d\omega_i+
  \frac13\epsilon_{ijk}\omega^i\wedge\omega^j\wedge\omega^k-\omega^i\wedge J_i \right),
\end{equation}
where $J$ is a two-form that corresponds to a source, whose explicit form will
remain unspecified for the time being.

The first two terms in the functional (\ref{CS}) can be recognized as the
Chern-Simons form associated to the Pontryagin invariant for $\MS\MO(3)$
\cite{Z,HZ}, which can also be seen as the  Euclidean continuation of the
so-called ``exotic" action for gravity in 2+1 dimensions \cite{Witten'88}. The
last term in the right hand side of Eq.(\ref{CS}) yields the interaction
energy between a connection and an external source, which generalizes the
minimal coupling between electric charges and the electromagnetic potential
\cite{Z09,MZ}.

The equilibrium configurations satisfy the conditions that make this functional
stationary under small variations, that is, they satisfy the equations
\begin{equation}                                                  \label{uvbxvr}
  R_{\mu\nu}{}^k=J_{\mu\nu}{}^k,
\end{equation}
where $J_{\mu\nu}{}^k$ are the components of the source term for the $\MS\MO(3)$
connection.

The interesting feature of both terms in this functional is that they are
invariant under gauge transformations
$\om^{ij}\mapsto\om'^{ij} =\om^{ij}+D\theta^{ij}$ up to a divergence term (see,
e.g., \cite{HZ}). Indeed, it can be shown that under a local gauge
transformation the Chern--Simons 3-form changes by an exact form, leaving the
field equations(\ref{uvbxvr}) unchanged. Additionally, the requirement that the
interaction term also changes by an exact form further demands that the current
$J$ describe a covariantly conserved source, $DJ=0$.
\section{Geometric analysis of a linear disclination}
We consider one linear disclination with the core along the line $q^\mu(t)$,
where $t\in\MR$ is a parameter along the core of disclination. Write the
invariant interaction term as
\begin{equation}                                                  \label{ujdght}
  S_\text{int}:=\int dq^\mu\om_{\mu i}J^i=\int dt\,\dot q^\mu\om_{\mu i} J^i.
\end{equation}
Suppose disclination is located in such a way that $\dot q^3\ne0$ everywhere.
To vary this action with respect to $\MS\MO(3)$ connection, we insert three
dimensional $\dl$-function:
\begin{equation*}
  S_\text{int}=\int dt d^3x \dot q^\mu\om_{\mu i}J^i\dl^3(x-q)=
  \int d^3x\frac{\dot q^\mu}{\dot q^3}\om_{\mu i}J^i\dl^2(x-q),
\end{equation*}
where we performed integration over $t$ using $\dl\big(x^3-q^3(t)\big)$ and
$\dl^2(x-q):=\dl(x^1-q^1)\dl(x^2-q^2)$ denotes the two dimensional
$\dl$-function on the $x^1,x^2$ plane. Then the variation is
\begin{equation}                                                  \label{ubcvdf}
  \frac{\dl S_\text{int}}{\dl\om_{\mu i}}=\frac{\dot q^\mu}{\dot q^3} J^i
  \dl^2(x-q).
\end{equation}

Consider equations (\ref{uvbxvr}) on topologically trivial manifold
$\MM\approx\MR^3$ with Cartesian coordinates $x^1=x$, $x^2=y$, and $x^3$.
Suppose disclination is located along $x^3$ axis, i.e.\ $q^1=q^2=0$ and
$q^3=t$. We are looking for solutions of equations (\ref{uvbxvr}) which are
invariant under translations along $x^3$ axis, and describe rotations only in
the $x,y$ plane. In this case $\MS\MO(3)$ connection has only two nontrivial
components:
\begin{equation}                                                  \label{uyertn}
  \om_x{}^3 \qquad\text{and}\qquad\om_y{}^3,
\end{equation}
which depend on a point on the plane $(x,y)\in\MR^2=\MC$. To find a solution, we
introduce complex coordinates
\begin{equation*}
  z:=x+iy,\qquad\bar z:=x-iy.
\end{equation*}
Then we have one complex valued component of $\MS\MO(3)$ connection
\begin{equation}                                                  \label{unvbgf}
\begin{split}
  \om_z{}^3&=\frac12\om_x{}^3-\frac i2\om_y{}^3,\\
  \om_{\bar z}{}^3&=\frac12\om_x{}^3+\frac i2\om_y{}^3
\end{split} \qquad \Leftrightarrow \qquad
\begin{split}
  \om_x{}^3&=\om_z{}^3+\om_{\bar z}{}^3,\\
  \om_y{}^3&=i\om_z{}^3-i\om_{\bar z}{}^3.
\end{split}
\end{equation}
The curvature tensor has only one linearly independent complex component
\begin{equation}                                                  \label{ubcvdr}
  R_{z\bar z}{}^3=2(\pl_z\om_{\bar z}{}^3-\pl_{\bar z}\om_z{}^3).
\end{equation}
The complex conjugate is
\begin{equation}                                                  \label{ubcvai}
  \overline{R_{z\bar z}{}^3}=2(\pl_{\bar z}\om_z{}^3-\pl_z\om_{\bar z}{}^3)
  =-R_{z\bar z}{}^3=R_{\bar z z}{}^3.
\end{equation}

If only two components of $\MS\MO(3)$ connection (\ref{uyertn}) differ from zero
then quadratic term in curvature (\ref{ushtde}) equals identically zero, and we
can consider $\dl$-function sources because equilibrium equations (\ref{uvbxvr})
become linear. Let us now fix the source term
\begin{equation}                                                  \label{ubdyui}
  R_{z\bar z}{}^3=4\pi i A\dl(z),\qquad A\in\MR,
\end{equation}
where $\dl(z)$ is the two-dimensional $\dl$-function on a complex plane. It is
clearly circularly symmetric.

Solution of Eq.(\ref{ubdyui}) describes the new kind of geometric defect. If
this equation were considered as the second order equation for the metric, then
the solution would describe conical singularity in metric on the $x,y$ plane.
In this case, the solution corresponds to the wedge dislocation in the geometric
theory of defects \cite{KatVol92}. Now the situation is different. We consider
this equation as the first order equation for the $\MS\MO(3)$ connection and
show that it describes defect in the spin structure, disclination, the metric
being Euclidean.

The solution to equation (\ref{ubdyui}) is
\begin{equation}                                                  \label{uxbvpo}
  \om_z{}^3=-\frac{iA}z,\qquad\om_{\bar z}{}^3=\frac{iA}{\bar z}.
\end{equation}
To check this one has to use well known equalities (see, i.e.\ \cite{Vladim71}):
\begin{equation}                                                  \label{ubvcfr}
  \pl_z\frac1{\bar z}=\pi\dl(z)\qquad\Leftrightarrow\qquad \pl_{\bar z}\frac1z
  =\pi\dl(z).
\end{equation}

The corresponding real components are
\begin{equation}                                                  \label{unbcyg}
  \om_x{}^3=-\frac{2Ay}{x^2+y^2},\qquad \om_y{}^3=\frac{2Ax}{x^2+y^2}.
\end{equation}

Outside the $x^3$ axis the curvature is flat and therefore given by partial
derivative of some function. In the geometric theory of defects, this function
is the rotational angle field $\theta(x,y)$ on the plane which describes
rotation of the unit vector field (spin structure) in media and must satisfy the
system of partial differential equations:
\begin{equation}                                                  \label{uxncrt}
  \pl_x\theta=-\frac{2Ay}{x^2+y^2},\qquad \pl_y\theta=\frac{2Ax}{x^2+y^2}.
\end{equation}
The integrability conditions for this system of equations
\begin{equation*}
  \pl_{xy}\theta=\pl_{yx}\theta
\end{equation*}
are fulfilled, and one can easily write a general solution
\begin{equation}                                                  \label{usgwed}
  \theta=-2A\arctan \frac xy+C,\qquad C=\const.
\end{equation}
Let us fix the integration constant $C:=\pi A$. Then the solution is
\begin{equation}                                                  \label{unbvtr}
  \tan \frac \theta{2A}=\frac yx=\tan\vf,
\end{equation}
where $\vf$ is the usual polar angle on the plane $(x,y)\in\MR^2$. If we go
along a closed loop $C$ around $x^3$ axis, then the polar angle varies from 0 to
$2\pi$. To make the rotational angle field $\theta(x,y)$ well defined, we must
impose the quantisation condition
\begin{equation}                                                  \label{ubcvdt}
  A=\frac n2,\qquad n\in\MZ.
\end{equation}

So, the rotational angle field is
\begin{equation}                                                  \label{uvxcsr}
  \theta=n\vf,
\end{equation}
where $\vf$ is the usual polar angle in the $x,y$ plane. It is defined
everywhere in the plane with the cut along half line, say, $y=0$, $x\ge 0$. The
corresponding $\MS\MO(3)$ connection is
\begin{equation*}
\begin{split}
  \om_x{}^{12}&=-\frac{ny}{x^2+y^2}=-n\sin\vf, \\
  \om_y{}^{12}&=~~\frac{nx}{x^2+y^2}=~~n\cos\vf.
\end{split}
\end{equation*}
It is well defined on the whole $x,y$ plane except the origin of the coordinate
system where its curl has $\dl$-function singularity (\ref{ubdyui}). We see that
the $\MS\MO(3)$ connection has much better behaviour then the corresponding
angular momentum field as it should be in the geometric theory of defects.

So, along a closed loop $C$ around $x^3$ axis, the rotational angle field
changes from 0 to $2\pi n$, where $|2\pi n|=|\Om|$ is the modulus of the Frank
vector. This is precisely the linear disclination in the spin structure of
media, the axis of disclination coinciding with the $x^3$ axis. For $n=0$
disclination is absent. This case should be treated separately: for $A=0$, there
must be $\theta=0$ as the consequence of Eq.(\ref{usgwed}). Two simplest
examples of linear disclinations for $n=1$ and $n=2$ are shown in
Fig.\ref{fdiscl} where we draw the distribution of unit vector field in the
$x,y$ plane.
\section{Conclusion}
We have shown that the Chern--Simons term in the geometric theory of defects is
well suited for description of linear disclinations. It produces nontrivial
Riemann--Cartan geometry in $\MR^3$. The metric is the flat Euclidean metric but
the $\MS\MO(3)$ connection is nontrivial. Curvature of the $\MS\MO(3)$
connection has a $\dl$-function singularity along $x^3$ axis. The singularity is
a new one. It is not a conical singularity in the $x,y$ plain (metric is
Euclidean) but the singularity is in the $\MS\MO(3)$ connection.

We compute explicitly the corresponding rotational angle field and show that it
describes the disclination defect in the spin structure. It is the first example
of disclination described within the geometric theory of defects.

The obtained string-type solution may have importance in gravity and cosmology 
(see, i.e.\ [17--19]).
\nocite{Katana15A,Katana15B,Katana16B}

The author would like to thank the Centro de Estudios Cientificos, Valdivia,
Chile and J.~Zanelli for collaboration.

\end{document}